\documentstyle{mn}
\title[K-band polarimetry of high-redshift
radio galaxies]{K-band polarimetry of seven 
high-redshift radio galaxies} 
\author[G. Leyshon and Stephen A. Eales]{G. 
Leyshon$^{1,2}$ and Stephen A. Eales$^{1,3}$\\ 
$^1$Department of Physics and Astronomy,
      University of Wales Cardiff,
      P.O. Box 913,
      Cardiff CF2 3YB\\
{$^2$ \verb"G.Leyshon@astro.cf.ac.uk"}\hspace{8mm}   {$^3$ 
\verb"S.Eales@astro.cf.ac.uk"}\hspace{8mm}
}

\input{psfig.sty}

\newcommand{\tpm}{$\pm\ $}

\newcommand{\tbn}[1]{${^{#1}}$}

\newcommand{\ah}[1]{{\hat{a}}_{\mathsf{#1}}}
 
\newcommand{\lums}[1]{{\lambda}_{\rm{#1}}}
 
\newcommand{\Itt}[1]{{F}_{\rm{#1}}}
 
\newcommand{\Iq}[1]{{F}_{\rm{Q,#1}}}

\newcommand{\Ptt}[1]{{P}_{\rm{#1}}}
 
\newcommand{\Phm}[1]{{\phi}_{\rm{#1}}}

\newcommand{\Ptm}[1]{{\Pi}_{\rm{#1}}}

\newcommand{\Stt}[1]{{\sigma}_{\rm{#1}}}
  
\newcommand{\Stm}[1]{{\epsilon}_{\rm{#1}}}
  
\newcommand{\Pq}[1]{{P}_{\rm{Q,#1}}}
 
\newcommand{\PHI}[1]{{\Phi}_{\rm{#1}}}
 
\newcommand{\ml}[1]{{m}_{\mathsf{#1}}}

\newcommand{\bpt}{$\bullet$}

\newcommand{\ccen}[1]{\multicolumn{1}{c}{#1}}

\newcommand{\cduo}[1]{\multicolumn{2}{c}{#1}}

\newcommand{\rS}{r_{\it S}}

\begin{document}

\maketitle

\begin{abstract}

 We present the results of K-band imaging polarimetry of seven 3CR radio
galaxies with $0.7<z<1.2$. We find strong evidence for polarization in
three sources: 3C 22, 3C 41 and 3C 114. Of these, 3C 41 shows strong
evidence of having a quasar core whose infrared light is scattered by
dust. We also find some evidence for polarization in 3C 54 and in 3C 356.
The two pointlike sources (3C 22 and 3C 41) and the barely-elongated 3C 54 
appear to have of 
order ten per cent of their K-band flux contributed by scattered light 
from the active nucleus. We
conclude that scattered nuclear light can form a
significant component of the near-infrared light emitted by high-redshift
radio galaxies, and discuss models in which the scattering particles are
electrons and dust-grains. 

\end{abstract}


\section{Introduction}

 A popular model for high-redshift radio galaxies was proposed by Lilly \&
Longair \shortcite{Lilly+84a} a decade ago. Their study of the K-$z$
Hubble diagram for 3CR radio galaxies showed a tight correlation: the 
K-magnitudes
exhibit a constant dispersion up to $z \ga 1$, but there is evolution with
redshift such that galaxies at $z \sim 1$ are about 1 mag more luminous
than at $z = 0$. Such a result is explicable if the observed K-band 
light originates in stars, and radio galaxies undergo a
burst of star formation at a very early epoch, followed by passive
evolution. A similar model has been suggested for radio-quiet 
elliptical galaxies \cite{Eggen+62a}.

 This model for radio galaxies was challenged by two discoveries, which 
suggested that a 
significant part of the the optical light originated in the active 
nucleus rather than the host galaxy. It was found that optical
light from high-$z$ radio galaxies is usually aligned with the radio axis
\cite{Chambers+87a,McCarthy+87a} -- the so-called `alignment effect' --
and that the optical light is often polarized with its $\bmath{E}$-vector
oriented perpendicular to the radio axis
\cite{Alighieri+89a,Jannuzi+91a,Tadhunter+92a}. The polarization of the
optical light suggests that it is emitted in a restricted range of 
directions by the active nucleus, and has
been scattered into our line of sight, by either dust or electrons. 

  Lilly \& Longair's \shortcite{Lilly+84a} model may still be correct if
the optical light originates in the active core while the (observer's
frame) K-band light comes from stars in the host galaxy. One test of this
is whether the alignment effect extends to the near-infrared; Dunlop \&
Peacock \shortcite{Dunlop+93a} find a clear infrared/radio alignment
effect for a sample of 3CR radio galaxies -- consistent with the smaller 3C
sample of Rigler et al.\ \shortcite{Rigler+92a}, which suggested that an
infrared alignment effect was present, but weak. 

 Eales et al. \shortcite{Eales+97a} compared the infrared luminosity of
3CR galaxies with that of a sample of B2/6C galaxies (selected for their
lower radio intensities in approximately the same frequency band as the
3CR galaxies).  They found that B2/6C galaxies are fainter than 3C
galaxies at similar redshifts, by 0.6 mags in K-band at $z \sim 1$; their
sample also confirms Dunlop \& Peacock's \shortcite{Dunlop+93a} finding
that the infrared alignment effect is strongest in
sources with high radio luminosity. Eales et al. \shortcite{Eales+97a} 
argue, therefore, that part of
the infrared luminosity of radio galaxies arises from the active nucleus,
and so a selection effect will produce the strong alignment effect and
brighter K-band magnitudes observed in the most radio-luminous galaxies,
{\it viz.} 3C galaxies at high $z$. One possible explanation of this
result is that a significant fraction of a radio galaxy's infrared light
emerges from the active nucleus in a restricted cone, and enters our line
of sight after scattering by dust or electrons. 
 There is already other evidence that light from the active nucleus, if
not necessarily scattered, can dominate the K-band light from high-$z$
radio galaxies: Dunlop \& Peacock \shortcite{Dunlop+93a} noted that radio
galaxy 3C 22 appeared to be exceptionally round and bright, and
spectroscopic studies by Rawlings et al.\ \shortcite{Rawlings+95a}
confirmed that in the K-band, 3C 22 has properties more similar to those
of a quasar than of a radio galaxy. 

We have started a programme of near-infrared polarimetry for two reasons. 
Detection of infrared polarization would reinforce the evidence that
scattered non-stellar light makes a significant contribution to the
infrared in radio galaxies. Also, K-band polarimetry provides a valuable
complement to optical polarimetry: a long wavelength baseline allows a
more powerful test of whether the scattering centres are electrons, with a
wavelength-independent scattering cross section, or dust, which
preferentially scatters short wavelengths \cite{Cimatti+93a}. 

\section{Observations and reduction procedure}

\subsection{Our sample}

 We observed a sample of seven high-redshift 3C radio galaxies at redshifts
$0.7<z<1.2$. These form a representative sample of the different 
morphologies present in this redshift band: 3C 22 and 3C 41 are bright and 
appear pointlike; 3C 114 and 3C 356 display complex knotted morphologies 
with large scale alignments between the K-band structure and the radio axis; 
3C 54, 3C 65 and 3C 441 are faint sources with some indication of K-band 
structure. Of these three faint sources, 3C 54 displays an alignment 
between the K-band morphology and the radio axis \cite{Dunlop+93a}, 3C 65 
shows no preferred direction in its H-band structure \cite{Rigler+94a}, 
and 3C 441 has a broad-band optical polarization which is 
roughly perpendicular to its radio structure \cite{Tadhunter+92a}.
Table \ref{obstab} lists the sources observed with their redshifts and the 
rest-frame wavelength of the observed light.

\begin{table}
\caption{Redshifts and integration times for our sample of radio 
galaxies.} 
\label{obstab}
\begin{tabular}{llrrr}
\hline \\
Source & IAU form & \ccen{$z$} & $\lambda_r$($\mu$m) & $t_{\rm int}$(min) \\
3C 22 & 0048+509 & 0.936 & 1.14 & 72 \\
3C 41 & 0123+329 & 0.794 & 1.23 & 108 \\
3C 54 & 0152+435 & 0.827 & 1.20 & 135 \\
3C 65 & 0220+397 & 1.176 & 0.92 & 72 \\ 
3C 114& 0417+177 & 0.815 & 1.21 & 189 \\ 
3C 356& 1723+510 & 1.079 & 1.06 & 99 \\
3C 441& 2203+292 & 0.707 & 1.29 & 144\\
\hline
\end{tabular}

\medskip

Key: $z$: redshift; $\lambda_r$: rest-frame equivalent of 
observed-frame 2.2$\mu$m;
$t_{\rm int}$: total integration time summed over all waveplate settings.
\end{table}

\subsection{Instrumentation}

 We had the first run after the commissioning run of a new instrument,
IRPOL2, installed at UKIRT (the United Kingdom Infrared Telescope,
Hawaii). The IRPOL2 polarimeter consists of a rotatable half-wave plate
and a Wollaston prism, working in conjunction with the IRCAM3 InSb array
detector. We used IRCAM3 at the default pixel scale, 0.286 arcsec/pixel
\cite{Aspin-94a}, with a K-band filter. The Wollaston prism causes each
source in its field of view to appear as a pair of superimposed images
with orthogonal polarizations, separated by $-0.93$ pixels in right
ascension and $+69.08$ pixels in declination \cite{Aspin-95a}. 

 The design of the instrument is such that when the waveplate is set to
its 0\degr\ reference position, an object totally linearly polarised with
its electric vector at 83\degr\ (i.e.  celestially East of North) would
appear only in the upper (Northern) image, and an object totally linearly
polarised at -7\degr\ would only appear in the lower image. There are four
standard offset positions for the waveplate: 0\degr, 22.5\degr, 45\degr\
and 67.5\degr. The IRPOL2 system has negligible instrumental polarization
\cite{hpc,Chrysostomou-96a}. We took data on the nights of 1995 August 25,
26, and 27. 

\subsection{Observing procedure}

\begin{figure}
\psfig{file=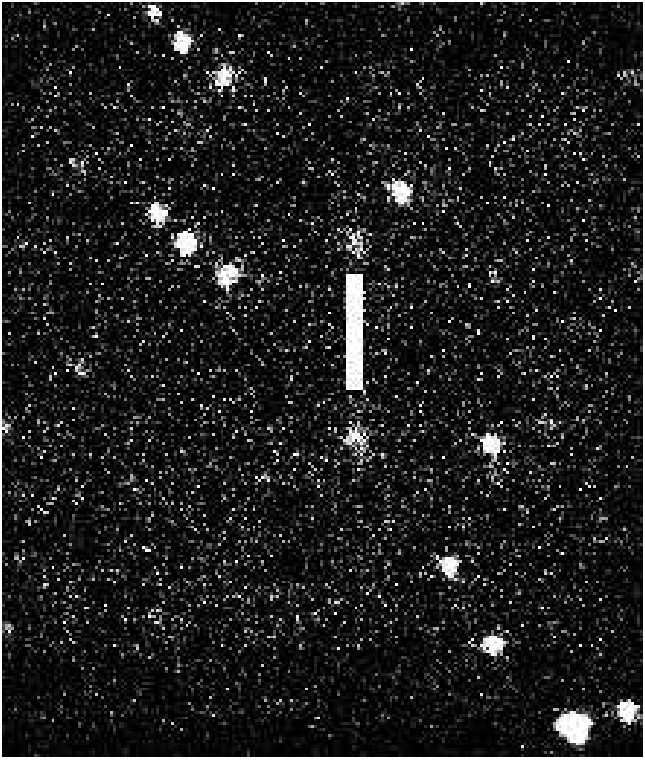,width=85mm}
\caption{K-band image of 3C 54 (at either end of the bar) and surrounding 
field. (North is at the 
top, East at the left. The image was composed by mosaicing a series of 
nine 60-second exposures taken with the waveplate offset at 22.5\degr.)} 
\label{54image} \end{figure}

 For each target object, we took a `mosaic' of nine images with the
waveplate at the 0\degr\ offset. One image consisted of a 60 second
exposure (the sum of six ten-second co-adds), and the mosaic was built up
by taking one image with the target close to the centre, and eight images
with the frame systematically offset from the first by $\pm 28$ pixels (8 
arcsec) horizontally and/or vertically. The same source was then similarly
observed with the waveplate at the 22.5\degr, 45\degr\ and 67.5\degr\
offsets, completing one cycle of observations;  hence one such cycle took
36 minutes of integration time. Between two and five observation cycles
were performed over the three nights for each target; the total integration
time for each target is given in Table \ref{obstab}. The times quoted are
not exact multiples of 36, as in some cases, mosaics were corrupted by
problems with the windblind, and excluded from our analysis. An example of a 
mosaic, 3C 54 and its
surrounding field, observed with the waveplate at 22.5\degr, is shown in
Figure \ref{54image}. 

\subsection{Data reduction}

 We reduced the data by marking bad pixels, subtracting dark frames, and
flat-fielding. Flat-fielding frames were obtained by median-combining the
nine images of each mosaic without aligning them, and normalizing the
resulting image by its mean pixel value. In order to align each set of
nine images, we chose a star present on each frame, and measured its
position with the {\sc apphot.center} routine in the {\sc iraf} package.
Using these positions, the nine images were melded into a single mosaic 
image. 

 We performed photometry on the pairs of images of field stars and of radio
galaxies in each mosaic using {\sc apphot.phot} from the {\sc iraf}
package. For each target, we chose one mosaic arbitrarily to determine 
the best aperture.
Using this mosaic, we performed photometry on the two images of the source
at a series of radii increasing in unit pixel steps. The measured
magnitude in each aperture decreases as the light included in the aperture
increases; we found the first pair $(r,r+1)$ of radii where the change in
magnitude was less than half the measured error on the magnitude. We then
used the photometry at the next radius, $(r+2)$. In this way, we hope to
include all the source light but as little as possible of the surrounding
sky. Where the chosen apertures differed for the two images of the source,
we adopted the larger. The chosen aperture was then used on all the
mosaics containing the target to obtain a set of intensities at the
corresponding waveplate positions.

 The {\sc apphot.phot} routine corrects for the sky brightness by
measuring the modal light intensity in an annulus around the target; the
position of the annulus was chosen in each case such that the outer radius
did not extend to the nearest neighbouring object, and the inner radius
was normally set one pixel greater than the photometry aperture. (Where we
attempted to perform photometry on a knot within a larger structure, the
inner radius of the annulus was set sufficiently large to exclude all the 
knots comprising the object.) This procedure results in 
pairs of photometric intensities for each target. 

\subsection{Confidence intervals for polarization}

 Using the photometric intensities from all the observing cycles, we
pooled the data and calculated the intensity-normalised Stokes Parameters,
$q$ and $u$, following the method of Clarke et al. \shortcite{Clarke+83a}.
The $q$ and $u$ (as quoted in Table \ref{polvals}) are in the instrumental
reference frame, such that a positive $q$ and zero $u$ would indicate
polarization at 83\degr\ East of North: $q$ is defined by subtracting the
lower beam intensity from the upper with the waveplate at the 0\degr\
reference position, and $u$ similarly with the waveplate at 22.5\degr. The
Stokes Parameters were then used to derive a measured degree of
polarization, $p= \sqrt{q^2 + u^2}$. 
 
Since $p$ is positive in
the presence of noise even for an unpolarised source, there has been some
debate about the best method of obtaining an `unbiased' estimate of the
polarization \cite{Simmons+85a,Leyshon-97a}.  Following best practice, we
assume that the Stokes Parameters $q$ and $u$ are normally distributed,
and that the errors on the two parameters are comparable. Under these
assumptions, it can be shown \cite{Wardle+74a,Vinokur-65a} that, if the
underlying polarization of a source is $p_0$ and the error on a
measurement is $\sigma$, then the probability distribution function for
the measured polarization values $p$ will be a Rice distribution. Defining
normalised variables $a=p_0/\sigma$ and $m=p/\sigma$, we can write this
distribution as
\begin{equation}
\label{rice}
 F(m,a) = m.\exp \left[ \frac{-(a^2+m^2)}{2} \right] .I_0(ma)
\end{equation}
where $m \gid 0$ and $I_0$ is the modified Bessel function.
It follows, trivially, that the integral
\begin{equation}
\label{cdef}
\int_{m=0}^{m=m_C(a)} F(m,a).dm = C
\end{equation}
for $0 \lid C \lid 1$ represents a confidence interval $(0,m_C)$ 
such that there is a chance $C$ of a measurement of the polarization lying 
within these bounds.

\begin{figure}
\psfig{file=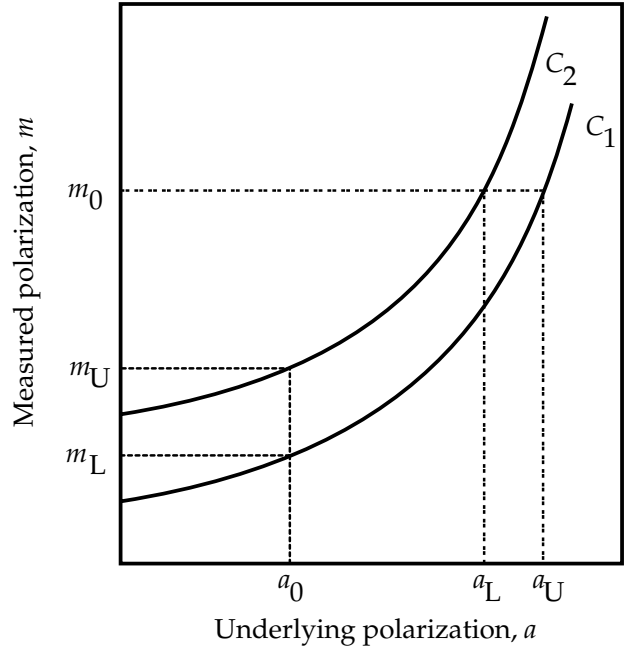,width=85mm}
\caption{Illustration of contours in the $a-m$ plane. See text.}
\label{contplot} \end{figure}

 We need to invert the probability distribution to estimate $a$ given 
our measured $m$. Following the method of Mood, Graybill \& Boes 
\shortcite{Mood+74a} we can solve Equation \ref{cdef} 
for $m_C$ for given values of $C$ and $a$, and so plot contours of constant 
$C$ on the $a$-$m$ plane, as illustrated in Figure \ref{contplot}. Clearly 
two contours $C_1$, $C_2$ allow us to 
read off a $C_2 - C_1$ confidence interval $(m_{\rm L},m_{\rm U})$ for given 
$a_0$. Mood et al. \shortcite[Ch.\ VIII, \S 4.2]{Mood+74a} show that the 
same contours also provide a $C_2 - C_1$ confidence interval for the 
underlying polarization $(a_{\rm L},a_{\rm U})$ for a given measurement 
$m_0$. 

 As well as providing a general interval $(a_{\rm L},a_{\rm U})$, the 
contours can be used to determine the probability that the 
underlying polarization is non-zero. The probability that $a=0$ exactly 
is finite rather than
infinitesimal, since some sources may be unpolarised.
There will be some contour $C_m$,
which cuts the $m$-axis at our measured $m_0$, and so
divides the domain of $a$ into $a>0$ with confidence $C_m$, and $a=0$ 
with confidence $1-C_m$. This confidence can be found analytically, since 
Equation \ref{cdef} can be integrated at $a=0$; and it follows that the 
probability that $a>0$, given some measured value $m$, is
\begin{equation}
\label{propol}
C_m = 1 - \exp(-m^2/2).
\end{equation}

 Equation \ref{propol} can be used to determine the probability that a 
source is truly polarised, and the probabilities so derived for each 
source are listed in Table \ref{polvals}.

 For a given measurement $m_0$, it is also possible to use appropriate
pairs of contours to find a set of 67 per cent ($1\sigma$) or 95 per cent
($2\sigma$) confidence intervals for our best estimate, $\hat{a}$. 
For any given required confidence, there will be a continuous set of 
pairs of contours which could be chosen; 
following Simmons \& Stewart \shortcite{Simmons+85a} we choose the unique
interval whose width is minimized by the additional constraint
\begin{equation}
\label{widcon}
F(m_{\rm L},a)=F(m_{\rm U},a).
\end{equation}
The errors on the polarizations quoted in Table \ref{polvals} are the 
$1\sigma$ errors derived by this method.

\subsection{Point estimates for polarization}

 To obtain a point estimate of the polarization, we again follow
Simmons \& Stewart \shortcite{Simmons+85a}, who have tested various
estimators $\ah{X}$ for bias. They find that when $a \la 0.7$, the best
estimator is the `Maximum Likelihood Estimator', $\ah{ML}$; 
when $a \ga 0.7$, the best estimator, $\ah{WK}$, is that traditionally 
used by radio
astronomers, e.g.\ Wardle \& Kronberg \shortcite{Wardle+74a}. It can be 
shown that $\ah{ML} < \ah{WK}$\ $\forall m$ \cite{Simmons+85a}.

 We do not know, {\it a priori}, whether $a$ will be greater or 
less than 0.7, and hence which estimator to use. We distinguish three cases:
(i) both estimators indicate $\ah{X} \la 0.7$, so we must use $\ah{ML}$;
(ii) both estimators indicate $\ah{X} \ga 0.7$, so we must use $\ah{WK}$; 
or (iii) in the region of $\ah{X} \sim 0.7$, each estimator will indicate 
that it should be preferred over the other. We therefore wish to 
construct an estimator $\hat{a}$ which is continuous with $m$ and takes 
these cases into account. This can be done using the inequality $\ah{ML} 
< \ah{WK}$\ $\forall m$.

 Case (i): 
Let $\ml{WKmin}$ be the measured value of polarization which would cause an 
estimate $\ah{ML} < \ah{WK} = 0.6$. Clearly if $m<\ml{WKmin}$ then 
$\hat{a} < 0.7$ and the Maximum Likelihood estimator is certainly the most
appropriate. We find that $\ml{WKmin} = 1.0982$ 
\cite{Leyshon-97a} and in this case,
the Maximum Likelihood estimator will in fact be zero.

 Case (ii): 
 Let $\ml{MLmax}$ be the measured value of polarization which 
would cause an
estimate $0.8 = \ah{ML} < \ah{WK}$. In this case, Wardle \& Kronberg's 
\shortcite{Wardle+74a} estimator will clearly be the
most appropriate. We find that $\ml{MLmax} = 1.5347$ \cite{Leyshon-97a}.

 Case (iii): Between these two extremes, if our measured polarization 
satisfies $\ml{WKmin} < m < \ml{MLmax}$, we can interpolate such that our 
best estimate is
\begin{equation}
\label{interpa}
\hat{a} = \frac{m-\ml{WKmin}}{\ml{MLmax}-\ml{WKmin}}.\ah{ML} +
\frac{\ml{MLmax}-m}{\ml{MLmax}-\ml{WKmin}}.\ah{WK}.
\end{equation}
 
 Thus we can obtain both a `debiased' point estimate and a confidence 
interval for $a$, and hence for the degree of polarization $a\sigma$.

\subsection{Estimating the angle of polarization}

 The position angle, $\phi = 83\degr + \phi_{m}$, of the 
$\bmath{E}$-vector is obtained by calculating
\begin{equation}
\label{redf}
2\phi_{m} = \theta = \tan^{-1}(u/q),
\end{equation} and noting the appropriate quadrant from the signs of $q$ 
and $u$. There are two methods of obtaining the error on this angle. We can 
propagate the errors on $q$ and $u$ through Equation \ref{redf}, yielding 
an asymmetric error; or the symmetric 67 per cent 
confidence interval may be constructed, using the best estimate of $a$ to 
calculate the relevant distribution function 
\cite{Wardle+74a,Vinokur-65a}. We calculated error bars by both methods, 
and selected the greater error in each case to quote in Table \ref{polvals}.

\subsection{Extinction corrections}

 The interstellar medium may impose some linear polarization on light 
transmitted through it. Empirically, the polarization $p$ depends on 
wavelength, such that
\begin{equation}
\label{semper}
p/p_{\rm max} = \exp [-1.7 \lums{max} \ln^{2}(\lums{max}/\lambda)] 
\end{equation}
where $\lums{max}$ is the wavelength (explicitly in micrometres) at which 
the polarization 
peaks, usually around $0.5 \mu$m, and empirically in the range 
0.3--0.8 $\mu$m \cite{Serkowski+75a,Wilking+80a}. The relationship is 
found to hold up 
to around $2\mu m$, and is adequate for our purposes \cite{Martin+90a}.

 In general, $p_{\rm max}$ for a given set of galactic co-ordinates is 
not known. But suppose we take the ratio of polarizations in two 
wavebands, V and K, and rearrange:
\begin{equation}
\label{sratund}
p_{\rm K} = p_{\rm V} \exp \left\{ -3.4 \lums{max} \left[ \ln 
\left( \frac{\lums{max}}{\sqrt{\lums{K}\lums{V}}}\right) \ln 
\left(\frac{\lums{V}}{\lums{K}}\right) \right] \right\}. \end{equation}
Hence $p_{\rm K} = c.p_{\rm V}$ where $c$ depends on $\lums{max}$ but for 
$0.3<\lums{max}<0.8$ we find $0.15<c<0.30$.

 An empirical upper limit for $p_{\rm V}$ (expressed as a percentage) is 
given by Schmidt--Kaler 
\shortcite{SchmidtKaler-58a} as $p_{\rm V} \lid 9E_{\rm B-V}$ and 
typically, $p_{\rm V} = 4.5E_{\rm B-V}$. It is well established 
\cite{Savage+79a,Koorneef-83a,Rieke+85a} that the ratio of total to 
selective extinction is $A_{\rm V}/E_{\rm B-V} \sim 3$; and so we can 
use the extinctions $A_{\rm B}$ \cite[figures obtained from the {\sc ned} 
database]{Burstein+82a} to obtain $E_{\rm B-V} = A_{\rm B} - A_{\rm V} = 
A_{\rm B}/4$.

 Taking the maximum values, $c=0.3$ and $p_{\rm V} \lid 9E_{\rm B-V}$, 
we find an upper limit for infrared polarization $p_{\rm K} \lid 
0.7A_{\rm B}$. The {\sc ned} values for $A_{\rm B}$ and the corresponding 
upper limits on $p_{\rm K}$ are given in Table \ref{extab}.

\begin{table}
\caption{Total B-band extinctions and upper limits on K-band interstellar 
polarization for our sample.}
\label{extab}
\begin{tabular}{lrr}
\hline \\
Source & $A_{\rm B}$ & $p_{\rm K,max}$ \\
3C 22 & 1.09 & 0.76 \\
3C 41 & 0.17 & 0.12 \\
3C 54 & 0.37 & 0.26 \\
3C 65 & 0.16 & 0.11 \\ 
3C 114& 1.26 & 0.88 \\ 
3C 356& 0.10 & 0.07 \\
3C 441& 0.34 & 0.24 \\

\hline
\end{tabular}

\medskip

Key: $A_{\rm B}$: Blue-band extinction (magnitudes), from {\sc ned}, 
derived from Burstein \& Heiles \shortcite{Burstein+82a}; $p_{\rm 
K,max}$: maximum interstellar contribution to K-band polarization (per 
cent) as described in the text.
\end{table}

\section{Results}

\subsection{Have we detected polarization?}

 Equation \ref{propol} allows us to quantify the probability that a given
object is polarised. The probabilities of each object being polarised are 
listed in Table \ref{polvals}. Three of our seven sources have a 95
per cent or better probability of being polarised; and of these, 3C 22 and
3C 41 are polarised at the 3 per cent level, and 3C 114 at the 12 per cent
level. 

The number of prominent starlike objects (in addition to the
target) featuring on the mosaics varies between one and seven, depending on
the target. Where possible, we have performed polarimetry on the stars; out
of the 21 stars so observed, only one has a greater than 95 per cent 
probability of being polarised. This object was a bright starlike object 
on the mosaic containing 3C 114, but is only polarised at the 
0.7 per cent level, which is explicable by the interstellar medium (see 
Table \ref{extab}).

 Within the bin of sources having a probability 80-95 per cent of being
polarized, fall three further stars; of these, one is extremely faint, and
another appears to be polarised at only the 0.3 per cent level. The third 
falls on
the same mosaic as 3C 54, and appears to be polarised at the 5.6
$\pm$ 2.6 per cent level, with a 94 per cent chance of the polarization 
being genuine. This star, however, straddles the edge of three of the nine
component frames of the mosaics, so the validity of the result is called
into question.  Two of our sources also fall in the 80-95 per cent
probability bin: 3C 54 itself, polarised at the 6 per cent level, and 3C
356, at the 9 per cent level. 

 Given that 17 out of 21 stars, but only 2 out of 7 sources, have a 
probability of less than 80 per cent of being polarised at all, we feel 
confident of having detected polarization in three 3C sources, and 
possibly in a further two.

\begin{table*}
\begin{minipage}{165mm}
\caption{Observational results from polarimetry of 7 radio galaxies in 
K-band.}
\begin{tabular}{lrrrrrrrrrrrr}
\hline \\
Source & $q (\sigma_q) $(\%) & $u (\sigma_u)$(\%)
& $r$(\arcsec) & $P \pm \sigma_P$(\%) & $2\sigma$.UL & Prob &
$\theta$(\degr) & \cduo{$\sigma_\theta$(\degr)}\\

3C 22 & -1.3 (1.4) & -3.2 (1.4) & 2.6 &  3.27 $\pm$ 1.38 & - & 0.95 & 27.0 & 
-71.4 & +16.9 \\

 3C 41 & +0.8 (1.1) & -3.1 (1.1) & 2.3  & 3.09 $\pm$ 1.11 & - & 0.98 & 
45.4 & -14.3 & +78.6 \\

3C 54  & -1.4 (2.5) & -6.0 (2.5) & 4.0 & 5.90 $\pm$ 2.55 & - & 0.94 & 31.6 & 
-79.1 & +17.3 \\

3C 65 & -4.3 (4.2) & -1.2 (4.0) & 2.9 & 2.15 $\pm$ \tbn{L}4.49 & 9.7 & 
0.42 & 0.6 & \cduo{$\pm$71.4} \\

3C 114&+11.2 (3.1) & -4.1 (2.7) & 3.6 & 11.74 $\pm$ 3.02 & - & 0.99 & 72.9 & 
-20.6 & +33.9 \\

\bpt Knee& +3.0 (1.6) & +4.3 (1.8) & 1.7 & 5.11 $\pm$ 1.73 & - & 0.99 & 
110.6 & -63.2 & +13.5 \\

3C 356& -9.6 (4.9) & +3.5 (5.6) & 4.0 &9.04 $\pm$ 4.94 & 16 & 0.85 & 172.0 & 
\cduo{$\pm$22.1} \\

\bpt North&  -13.3 (7.9) & +4.6 (8.4) & 2.3 & 13.02 $\pm$ 7.99 & 41 & 0.62 & 
163.5 & \cduo{$\pm$25.0} \\ 

\bpt SE    &  -10.1 (9.4) &-18.9 (16.7) & 2.6 &  18.99 $\pm$ 15.31& 24& 
0.78 & 23.9&-54.2&+33.0 \\

3C 441&-1.7 (2.3) & -0.8 (2.3) & 3.4 & 1.04 $\pm$ \tbn{L}2.35 & 4.7 & 0.26 & 
6.2 & \cduo{$\pm$77.3} \\

\hline
\end{tabular}

\medskip

Key: $q \pm \sigma_q, u \pm \sigma_u$: normalized Stokes
Parameters (per cent) with respect to 83\degr\ E of N; r: radius of 
photometry aperture (arcseconds);
$P\pm\sigma_P$: percentage polarization (debiased) with 1$\sigma$ error
(\tbn{L} --- the $1\sigma$ lower limit
for polarization is zero); Prob: the probability that there is underlying 
polarization, given by Equation \ref{propol}; $2\sigma$.UL: $2\sigma$ upper 
limit (in per cent) for polarization in 
objects unlikely to be polarised; $\theta\pm\sigma_\theta$: Electric vector 
orientation E of N (\degr).

\label{polvals}
\end{minipage}
\end{table*}

\subsection{Individual objects}

\subsubsection*{3C 22}

3C 22 has a 95 per cent chance of truly being polarised, and the debiased 
polarization is $3.3\pm1.4$ per cent. We expect that no more than 0.8 per 
cent is due to the interstellar medium; most of the polarization is 
therefore intrinsic to the source. The orientation of the 
$\bmath{E}$-vector is ${+27}^{+17}_{-71} \degr$ East of North. Our K-band 
image shows no evidence for extended structure.

 The radio position angle of 3C 22 is 102\degr\ \cite{Schilizzi+82a}. The 
errors on the measured
polarization orientation are large, but the figures suggest that the true
direction is more likely to be perpendicular to the radio axis, than
parallel to it. As we have already seen, 3C 22 is suspected of being an
obscured quasar \cite{Dunlop+93a,Rawlings+95a}; according to data in 
Lilly \& Longair \shortcite{Lilly+84a}, its K-band magnitude ($15.67 \pm 
0.10$) its brighter than the mean K-$z$ relationship by about 0.9 mag.
 A perpendicular alignment of the
radio axis and $\bmath{E}$-vector orientation would be expected for light
from the active nucleus travelling along the radio axis and then 
scattered towards us by either dust or electrons.
 Jannuzi \shortcite{jpc} \cite{Elston+97a} 
has performed imaging polarimetry on 3C22 at shorter wavelengths, and 
reports $3\sigma$ upper limits in V and H of 5 per cent and 3 per cent 
respectively.

\subsubsection*{3C 41}

3C 41 has a 98 per cent chance of having an underlying polarization, 
which we measure to be 3.1 $\pm$ 1.1 per cent. Our upper limit for 
extinction-induced polarization is only 0.1 per cent, so we are confident 
of having detected intrinsic polarization in this object.
The orientation of the
$\bmath{E}$-vector is ${+45}^{+79}_{-14} \degr$ East of North. Our K-band
image shows no evidence for extended structure. Like 3C 22,
Lilly \& Longair`s \shortcite{Lilly+84a} data shows that, at K $= 15.95 
\pm 0.10$, 3C 41 is significantly brighter than the mean K-$z$ 
relationship, by about 0.6 mag.
 
 Jannuzi \shortcite{jpc} \cite{Elston+97a} 
have firm V and H band polarizations for this source:
at V, the polarization is 9.3 $\pm$ 2.3 per cent at 
+58\degr $\pm$ 7\degr; at H, the polarization is 6.6 $\pm$ 1.6 per cent at 
+57\degr $\pm$ 7\degr. The $\bmath{E}$-vector orientations in the three 
wavebands are consistent with one another. The radio position angle of 3C 
41 is 147\degr\ \cite{Longair-75a}, so we have a very good perpendicular 
alignment between the radio and polarization axes.

\subsubsection*{3C 54}

 The probability that 3C 54 is polarised is 94 per cent;
our measured value of polarization is 5.9 $\pm$ 2.6 per cent 
at ${+32}^{+17}_{-79} \degr$ East of North. Dust is not expected to 
contribute more than 0.3 per cent. We have no independent polarimetry for
this object. The position angle of its radio structure is 24\degr\ 
\cite{Longair-75a}; our measurement shows that the polarization 
orientation is more likely to be parallel to the radio axis, than 
perpendicular to it, but the latter case cannot be ruled out. 
Dunlop \& Peacock's \shortcite{Dunlop+93a} K-band contour map shows a 
slight extension to the south-west, which was also just discernible in 
our image. This would be roughly parallel with the radio axis.

\subsubsection*{3C 114}

\begin{figure}
\psfig{file=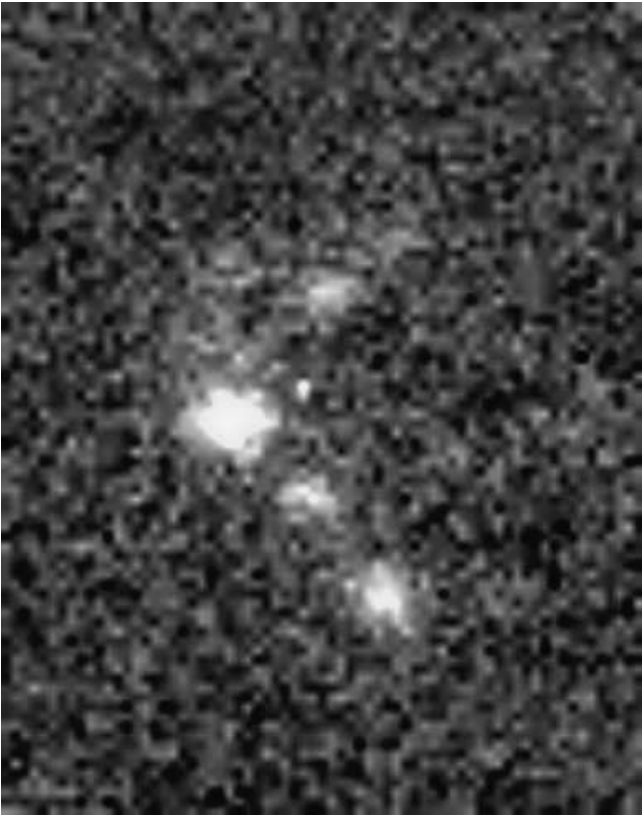,width=85mm}
\caption{K-band structure of 3C 114. (North is at the top, East is on 
the    
left.)} 
\label{114image}
\end{figure}

 An image of 3C 114 is shown in Figure \ref{114image}. It consists of at 
least four knots, where the brightest knot forms the knee of a
$\Gamma$-shaped structure. Only the knee knot proved bright enough to 
analyze on its own; Table \ref{polvals} includes results for both that knot 
and the structure as whole.
There is a probability in excess of 99 per cent that there is genuine 
polarization in both the knee knot and the overall structure. 

 We measured polarization in the knee of 5.1 $\pm$ 1.7 per cent, at 
${+111}^{+14}_{-63} \degr$ East of North. Overall, the whole object has a 
polarization of 11.7 $\pm$ 3.0 per cent at ${+73}^{+34}_{-20} \degr$. The 
extinction contribution could be as high as 0.9 per cent, but our 
detections of polarization are much higher than this, so the polarization 
appears to be intrinsic.

 The radio position angle of 3C 114 is 54\degr\ \cite{Strom+90a}.
Our measurements of the $\bmath{E}$-vector 
orientation suggest that an alignment parallel to the radio and optical 
axes is more likely than a perpendicular alignment, but the errors do not 
permit a more definite conclusion.

\subsubsection*{3C 356}

 We have no strong evidence for polarization in 3C 356, but we note that the 
structure has three lobes in its K-band image, at the North, the 
South-East and the South-West. Only the North and South-East 
knots were bright enough to permit polarimetry, as recorded in Table 
\ref{polvals}.

\section{Discussion}

\subsection{The fractional contribution of quasar light}   

 Radio galaxies are thought to have quasar nuclei at their cores, but to
be oriented such that no direct radiation from the core can reach us. 
Nearby radio galaxies are known to have the morphology of giant 
ellipticals; if this is also true at higher redshifts, the stellar 
populations of our seven sources will consist of mostly old, red, stars. 
The total light received from our sources will be a combination 
of starlight, light from the active nucleus scattered into our line of 
sight, and nebular continuum emission.

 We denote by $\PHI{W}$, the fraction of the total flux density,
$\Itt{W}$, in a given waveband, W, which originates in the active 
nucleus. We expect that in the visible wavebands, $\PHI{U,B,V}$ will be a 
significant fraction of unity. From our observations, we wish to 
determine whether $\PHI{K}$ is small, or whether a significant component 
of the stellar-dominated infrared also arises in the active nucleus. 

 Where the magnitude, W, has been measured in a given waveband, the total 
flux density can be calculated:
\begin{equation}
\label{totint}
\Itt{W}=\Itt{0}({\rm W}=0).10^{-0.4{\rm W}};
\end{equation}
if we denote the flux density scattered into our line of sight from the 
quasar core by $\Iq{W}$, then $\PHI{W} = \Iq{W}/\Itt{W}$. 

 The optical flux density of quasars can be modelled well by a power law
of the form $C.\nu^{-\alpha}$ where $\alpha$, the `spectral index', is of
order unity. Since the efficiency with which a given species of scattering 
centre 
scatters light may depend on wavelength, we denote that efficiency by 
$f_{\rm W}$, and the scattered quasar component can be expressed
\begin{equation}
\label{scatcomp}
\Iq{W}= C.f_{\rm W}.\nu^{-\alpha}.
\end{equation}

It will be computationally convenient to define an `unscaled model
light ratio', $\Phm{W}$, as
\begin{equation}
\label{phiunsc}
\Phm{W} =
\frac{f_{\rm W}.\nu^{-\alpha}}{\Itt{0}(W=0).10^{-0.4{\rm W}}};
\end{equation}
then the actual model light ratio will be $\PHI{W}=C.\Phm{W}$.
There is clearly an upper limit set on $C$ by the fact that $\PHI{W}$ 
may not exceed unity in any waveband; hence $C \leq 1/\Phm{W}$. To 
allow for errors in the observed magnitudes, $\Delta W$, the maximum 
permissible value of $C$ in a model can be taken to be
\begin{equation}
\label{conmax}
C_{\rm max} = {\rm min} \left[ \frac{\Itt{0}({\rm W}=0).10^{-0.4({\rm 
W}-\Delta{\rm W})}}{f_{\rm W}.\nu^{-\alpha}}\right], \forall\ {\rm W},
\end{equation}
where the value for $\alpha$ and the 
choice of scattering centre (hence of $f_{\rm W}$) depend on the model.

\subsection{The dilution law for polarization}   

 Following Manzini \& di Serego Alighieri \shortcite{Manzini+96a}
(hereafter MdSA) we assume that only the scattered component of the light
from radio galaxies is polarised. It can readily be shown that when
linearly polarised light is mixed with unpolarised light, the overall
degree of polarization is directly proportional to the fraction of the
total intensity contributed by the polarised component. Hence defining
$\Pq{W}$ as the intrinsic polarization produced by the scattering process,
and $\Ptt{W}$ as the observed polarization after dilution, it follows that
\begin{equation}
\label{fracflux}
\PHI{W} = \Iq{W}/\Itt{W} = \Ptt{W}/\Pq{W}.
\end{equation}

 If we know the restrictions on possible values of the intrinsic
polarization $\Pq{W}$, we can use our corresponding measurements of
$\Ptt{W}$ to put limits on $\PHI{W}$ for the measured sources. The 
appropriate restrictions depend on the nature of the scattering centres.

If the scattering centres are electrons \cite{Fabian-89a}, then Thomson
scattering will take place: the effects of the geometry and of the
wavelength can be treated independently. The spectral energy distribution
of the light scattered in a given direction is independent of the
scattering angle:  $\Pq{W}=\Ptt{Q}$ will be a constant. The degree of
polarization of the scattered light is given simply by 
\begin{equation}
\label{elecpol} 
\Ptt{Q} = \frac{1-\cos^2\chi}{1+\cos^2\chi}, 
\end{equation}
where $\chi$ is the scattering angle. For an AGN observed as a radio 
galaxy, we assume an orientation 45\degr\ $\leq \chi \leq$ 90\degr, whence 
$1/3 \leq \Ptt{Q} \leq 1\ \forall\ $W.

 The case where the scattering centres are dust grains has been modelled
recently (MdSA); the fraction of the light scattered by the dust, $f_{\rm
W}$, and the polarization of the scattered light, $\Pq{W}$, both depend
strongly on wavelength. The exact relationship depends critically on the
size distribution of the dust grains, and the amount of extinction they
introduce; MdSA provide a series of graphs for the variation of $f_{\rm W}$ 
and $\Pq{W}$ with rest-frame wavelength
0.1$\ \mu$m $< \lambda_r < 1.0\ \mu$m, corresponding to many different dust
grain compositions and size distributions. At the redshifts of the objects
in our sample, the light observed in the H and J bands originated at 
rest-frame wavelengths below 1.0\ $\mu$m, but the K-band light originated in 
the region 
1.0$\ \mu$m $< \lambda_r < 1.15\ \mu$m. To accommodate the K-band light 
within our models, we linearly
extrapolated MdSA's curves to $\lambda_r = 1.15\ \mu$m.

In models where the smallest dust grains have a radius not less than 40
nm, $\Pq{W}$ approaches zero twice: once at a (rest frame) wavelength
around 0.2 $\mu$m, and again at some wavelength between 0.1 and 0.7 $\mu$m
which depends critically on the dust grain size distribution. But in all
cases, $\Pq{W}$ extrapolated into the 1.0 $\mu$m $< \lambda_r < 1.15\ \mu$m
region gives 0.3 $< \Pq{K} <$ 0.5; the intrinsic polarization at V 
and H is lower. 

 For the five sources in which we have evidence of K-band polarization, we
can hence estimate $\PHI{K}$ under both electron and dust models. The
values are given in Table \ref{quasfrac}. 
For the dust models, we take $ 1/\Pq{W} = 2.5 \pm 0.5$; the error takes
into account all dust models, and the different redshift corrections for
the different galaxies, but assumes that the scattering angle is 90\degr.
If the scattering angle is less, we assume that less polarization occurs
(see MdSA, Figure 20), and hence $\PHI{K}$ will be greater than our estimate.
For the electron models, we multiply the observed polarization by $
1/\Pq{W} = 2 \pm 1$; this takes into account all possible orientation
effects. 

\begin{table}
\caption{Percentage of K-band light estimated to be arising from the active 
nucleus in the 5 polarized galaxies.}
\label{quasfrac}
\begin{tabular}{lccc}
\hline \\
Source & $\Ptt{K}$ & ${\PHI{K}}_{\it e}$ & ${\PHI{K}}_{\it d}$ \\
3C 22 & 3.3 \tpm 1.4 & 7 \tpm 4 & 8 \tpm 4 \\
3C 41 & 3.1 \tpm 1.1 & 6 \tpm 4 & 8 \tpm 3 \\
3C 54 & 5.9 \tpm 2.6 & 12 \tpm 8 & 15 \tpm 7 \\
3C 114& 11.7 \tpm 3.0 & 23 \tpm 13 & 29 \tpm 10 \\ 
3C 356& 9 \tpm 5 & 18 \tpm 13 & 23 \tpm 13 \\

\hline
\end{tabular}

\medskip

Key: $\Ptt{K}$: measured K-band polarization; ${\PHI{K}}_{\it e}$: 
fraction of light from quasar according to electron model; ${\PHI{K}}_{\it 
d}$: fraction of light from quasar according to dust model. 
Depolarization from multiple scattering (both models), and shallower 
scattering angles (dust model only) will tend to increase $\PHI{K}$. The 
polarizations used for the compound objects 3C 114 and 3C 356 are those 
for the whole compounds.
\end{table}

 The effects of multiple scattering have been ignored for both models; 
multiple scattering tends to depolarise light, and so the true value of
$\PHI{K}$ under multiple scattering will again be greater than our
estimate. The only physical mechanism which could cause $\PHI{K}$ to be 
{\em lower} than our estimate, is polarization of light 
in transit by selective extinction; and as we have already seen 
(Table \ref{extab}), any such contribution to the polarization will 
be small.

\subsection{Modelling the polarization of radio galaxies}

Given a measurement of the magnitude of a radio galaxy, we can predict 
its polarization as a function of 
wavelength, up to a multiplicative constant. Rearranging Equation 
\ref{fracflux}, and employing our `unscaled model light ratio',
we first define an `unscaled model polarization' $\Ptm{W} = \Pq{W}.\Phm{W}$, 
and so express our modelled polarization as: 
\begin{equation}
\label{polmod}
\Ptt{W,modelled} = \Pq{W}.\PHI{W} = 
C.\Pq{W}.\Phm{W} = C.\Ptm{W}.
\end{equation}

To fit a dust scattering model, we can calculate $\Ptm{W}$ by obtaining 
$f_{\rm W}$ and $\Pq{W}$ from suitable 
curves in MdSA. For electron models, the wavelength-independent term 
$f_{\rm W}$ can be considered to have been
absorbed into the multiplicative constant, $C$, while $\Ptt{Q}$, also 
wavelength-independent, can be assumed to be its minimum value, 1/3. 
We cannot separately identify $C$ and $\Ptt{Q}$; the physical 
constraints on these constants are $1/3 \leq \Ptt{Q} \leq 1$ and $0 \leq C 
\leq C_{\rm max}$. If $\Ptt{Q}$ is greater than the assumed 1/3, then $C$ 
will be correspondingly smaller.

 Given a set of $N$ polarization measurements $\Ptt{W} \pm \Stt{W}$, and a 
corresponding set of unscaled model polarizations, $\Ptm{W} \pm \Stm{W}$, 
based on measured magnitudes, we can calculate the deviation of the fit:
\begin{equation}
\label{fitdev}
\delta = \sqrt{\frac{1}{N}.\sum_{W_{1}\ldots 
W_{N}}{\frac{(\Ptt{W}-C.\Ptm{W})^2}{\Stt{W}^2+(C.\Stm{W})^2}}}. 
\end{equation}
The best fit is that with the value of $C$ which minimizes $\delta$,
subject to the physical constraint $0\leq C\leq C_{\rm max}$.

\subsubsection*{A model for 3C 41}

 The source for which we had the most data was 3C 41, with polarimetry in 3 
bands: $\Ptt{V}=9.3$ \tpm 2.3 per cent; 
$\Ptt{H}=6.6$ \tpm 1.6 per cent; $\Ptt{K}=3.1$ \tpm 1.1 per cent.
We attempted to fit two models; an electron model and
 a typical dust model with a minimum grain radius of 80 nm. 
To fit the models to the observed polarizations, we tested a discrete series 
of possible spectral indices, $-0.5 \leq \alpha \leq 2$, with a 
step size of $1/3$.
For each value of $\alpha$ we calculated the `unscaled polarizations' 
$\Ptm{V}\pm\Stm{V},\Ptm{H}\pm\Stm{H},\Ptm{K}\pm\Stm{K}$. We then
iteratively determined the best fit value of $C$ for each $\alpha$, 
and took as our overall best fit that 
combination of $\alpha$ and $C$ which gave the lowest $\delta$.

Magnitudes for 3C 41 were available in 5 bands: J, H and K
\cite{Lilly+84a} and the narrow filters g and r$_S$ \cite{dpc}. The H and
K values yielded direct estimates of the corresponding `unscaled
polarizations' $\Ptm{H}$ and $\Ptm{K}$; $\Ptm{V}$ was estimated by linear
interpolation between $\Ptm{g}$ and $\Ptm{\rS}$. 

\begin{figure}
\psfig{file=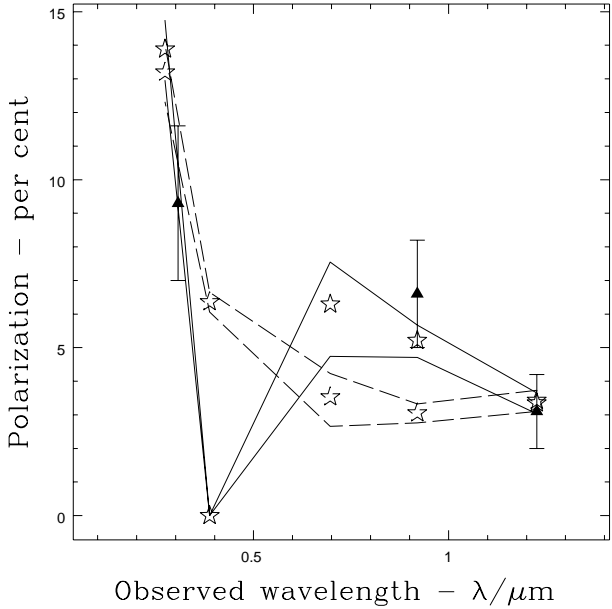,width=85mm}
\caption{Measured and best-fit model polarizations of 3C 41 as a function of 
rest-frame wavelength. Solid line: dust 
model, $\alpha = 1.167$; dashed line: electron model, $\alpha = 1.733$.} 
\label{pmodel41} \end{figure}

Figure \ref{pmodel41} shows the measured polarizations for 3C 41 as
triangles (\psfig{file=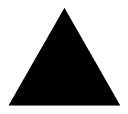,height=0.9em}) and the magnitude-based
polarization estimates, after best-fit scaling, as stars
(\psfig{file=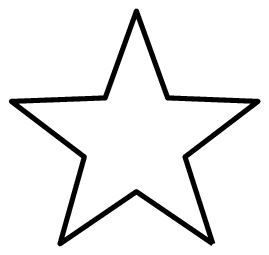,height=0.9em}). The lines give the error envelope on
the modelled polarization (based only on the errors on the magnitudes).
The solid lines correspond to the dust model, and the dashed lines to the
electron model. 

 As can be seen from Figure \ref{pmodel41}, the model curves lie below the
data point at H, but above those at K and V. The shape of the curve depends 
more
strongly on the measured magnitudes (and on $f_{\rm W}$ for dust) than on
the spectral index, and the models for all reasonable values of $\alpha$
will have a broadly similar shape; the best fit will necessarily pass
below the measured point at H, and above that at V. 

 Consider the electron model. Figure \ref{pmodel41} shows
us that the theoretical polarization curve for electron scattering is
concave with respect to the origin, whereas a curve through the three data
points would be convex; clearly it will not be possible to obtain a close
fit for the central (H-band) point. Fitting the
electron model curve, we found that the best fit occurred for $\alpha =
1.733$, with a deviation $\delta = 0.430$.
From Table \ref{quasfrac}, we have $\PHI{K}
= 6 \pm 4$ per cent for 3C 41. Multiplying the polarizations observed at H 
and 
V by $1/\Pq{W} = 2 \pm 1$, we predict $\PHI{H}=13 \pm 7$ per cent, and
$\PHI{V}=19 \pm 10$ per cent in these bands.

 The dust model chosen as typical from MdSA was that for a cloud of
spherical dust grains, with radii 250 nm $> a >$ 80 nm, with the number
density per unit dust mass following an $a^{-3.5}$ law. This model produced
a curve which fitted the data points very well. The best fit indicated
that the optimum spectral index was $\alpha = 1.167$, for which $\delta =
0.177$. 

 This dust model was also used to estimate the proportion of scattered
light at shorter wavelengths: $\PHI{K} = 8 \pm 4$ per cent, $\PHI{H} = 24
\pm 6$ per cent, and $\PHI{V} = 155 \pm 38$ per cent. MdSA's curve for 
polarization as
a function of rest-frame wavelength predicts a 6 per cent polarization in 
the observed
V-band, lower than the 9 per cent {\it after dilution} measured by Jannuzi
\shortcite{jpc} \cite{Elston+97a}. This is still consistent, within error
bars, as long as the true value of $\PHI{V}$ for 3C 41 is less than, but
very close to, unity;  the observed V-band corresponds to the
near-ultraviolet in the rest frame of 3C 41, and it is reasonable (MdSA)
to suppose that the scattered quasar light in that band could form in
excess of eighty per cent of the total light. 

We noted earlier that the shape of the dust model polarization curve
between 0.2 $\mu$m and 0.7 $\mu$m (rest frame) is very sensitive to the
choice of dust grain distribution. The particular dust model chosen
approaches zero polarization at a wavelength corresponding to the r$_S$
band when redshifted into our frame. This causes the `well' visible in the
model polarization curve (Figure \ref{pmodel41}), whose presence is 
essential for the dust model curve to fit the data points closely.

 We also considered other dust models from the selection given by MdSA. 
Of those which differed significantly from the `typical' one considered so
far, many of them will not predict polarizations in the observed V-band
which are sufficiently high to be reconciled with the observed 9 per cent;
and those which do, do not possess the deep well needed to fit the
polarimetry across the spectrum. We conclude, therefore, that
the best model for 3C 41 is that of an obscured quasar core with $\alpha
\sim 1.2$ beaming its optical radiation into a dust cloud, although an 
electron model cannot be ruled out within the error bars.

\subsubsection*{A model for 3C 22}

\begin{figure}
\psfig{file=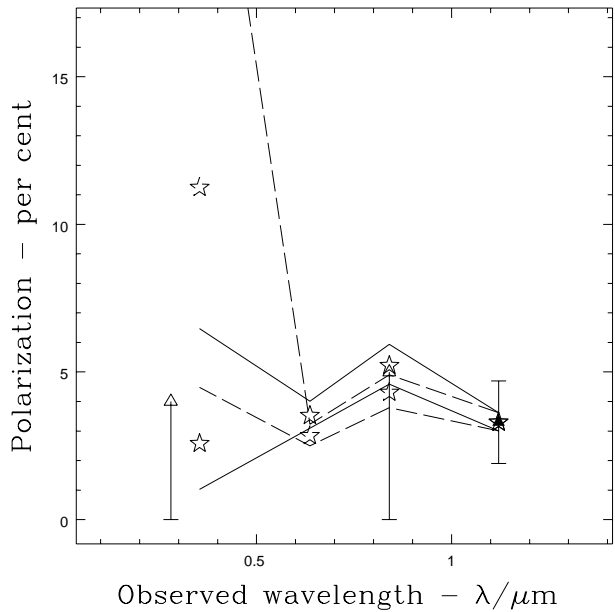,width=85mm}
\caption{Modelled ($\alpha = 1$) and measured polarizations of 3C 22 as a 
function of 
rest-frame wavelength. Solid line: dust model; dashed line: electron model.}
\label{pmodel22}
\end{figure}

For 3C 22, we have one firm polarimetry point (this paper) and two upper 
limits in V and H \cite{jpc} \cite{Elston+97a}; magnitudes were available 
in 4 bands including J, H and K \cite{Lilly+84a}, and a crude eye 
estimate in r \cite{Riley+80a}. The measurements and models are shown in 
Figure \ref{pmodel22}, with the same symbols as Figure \ref{pmodel41}; 
open triangles represent upper limits. We 
have taken $\alpha = 1$, and normalised the theoretical curves to the 
K-band data point.

Here the models are inconclusive. In the observed near-infrared, both 
models can easily fit, with some slight scaling, within the K-band 
measurement error bars; both models suggest that about 8 per cent of the 
K-band light arises in the active nucleus (Table \ref{quasfrac}). The 
error on the observed r-band magnitude is so 
large that both models are consistent with the observed V-band upper 
limit of polarization.

\section{Conclusions}
 
We have taken K-band polarimetry for seven 3CR radio galaxies, and found 
a diverse range of results. Out of our seven sources, 
two (3C 65 and 3C 441) display no evidence for polarization.
For the five sources which do display some evidence of polarization, we 
have estimated the fraction of observed K-band light which originated in 
the active nucleus.

 The recent finding \cite{Eales+97a} that 3C galaxies 
at $z \sim 1$ are 1.7 times as bright as the radio-weaker 6C/B2 galaxies 
in a similar sample requires that $\PHI{K} \ga 40$ per cent $(= 7/17)$ 
for 3C galaxies, if the scattering of nuclear nonstellar light is 
responsible for the brighter K-band magnitudes of 3C galaxies. Our results 
suggests that $\PHI{K}$ is somewhat lower. Nevertheless, our 
results support the hypothesis that radio galaxies consist of quasar 
nuclei embedded in giant elliptical galaxies.

 All of the galaxies which appear to be polarised have large errors on 
the orientation of the $\bmath{E}$-vector; hence any apparent 
alignment effects are suggestive rather than definitive. But with 
this caveat, we note that two sources (3C 54 and 114) display extended 
K-band structure, and have high polarizations oriented in roughly parallel 
alignment with the radio axis and extension of the optical structure -- 
i.e., in the opposite sense to the perpendicular alignment expected under a 
simple scattering model.

 Finally, 3C 22 and 3C 41 display significant polarization of around 3 per
cent, with a polarization alignment perpendicular to their radio axes;
both appear in K-band as pointlike objects. We suggest, therefore, that in
these objects, infrared light from a quasar core is being scattered into
our line of sight, and forms a significant part of the total K-band flux
received from these sources. 

Manzini \& di Serego Alighieri (MdSA) comment that given the range of
possible dust models, `the wavelength dependence of polarization is not
necessarily a discriminant between electron and dust scattering.' The
data available to us are insufficient to indicate whether the scattering
centres in these objects are electrons or dust; it is not possible to give
an unambiguous fit of the polarization curves with so few data points,
although 3C 41 does seem to fit a model (MdSA) with a minimum dust radius
of 80 nm, particularly well, and we suggest that it does indeed consist 
of a quasar obscured by dust. 

\subsection*{Acknowledgements}

 We thank Steve Rawlings for his encouragement in this project; Colin 
Aspin and Tim Carroll for help with making the observations; Buell
Jannuzi and Richard Elston for sharing their polarimetry results; Mark
Dickinson for some optical magnitudes; and Jim Hough, Chris Packham and 
Antonio Chrysostomou for
help with our polarimetry reduction. Simone Bianchi, Paul Alton, Bryn
Jones, Bob Thomson, Neal Jackson, Arjun Dey, Clive Tadhunter and Patrick
Leahy provided useful references. We also thank the anonymous referee for 
his constructive comments.

 This research has made use of the {\sc nasa/ipac} extragalactic database
({\sc ned}) which is operated by the Jet Propulsion Laboratory, CalTech, 
under contract with the National Aeronautics and Space Administration.

The United Kingdom Infrared Telescope is operated by the Joint Astronomy 
Centre on behalf of the U.\,K. Particle Physics and Astronomy Research 
Council. We thank the Department of Physical Sciences, University 
of Hertfordshire for providing IRPOL2 for the UKIRT. GL thanks {\sc 
pparc} for a postgraduate research student award.

\end{document}